\documentclass[a4paper, amsfonts, amssymb, amsmath,  showkeys, nofootinbib, twoside, superscriptaddress]{revtex4-1}
\usepackage[english]{babel}
\usepackage[utf8]{inputenc}

\usepackage[colorinlistoftodos, color=green!40, prependcaption]{todonotes}

\usepackage{amsthm}
\usepackage{mathtools}
\usepackage{physics}
\usepackage{xcolor}
\usepackage{graphicx}
\usepackage[left=23mm,right=13mm,top=35mm,columnsep=15pt]{geometry} 
\usepackage{adjustbox}
\usepackage{placeins}
\usepackage[T1]{fontenc}
\usepackage{csquotes}

\usepackage{natbib} 
\usepackage[version=4]{mhchem}
\usepackage{geometry}
\usepackage{graphicx} 

\usepackage{epstopdf}

\usepackage{tikz}
\usepackage{multirow}

\DeclareRobustCommand\sampleline[1]{%
  \tikz\draw[#1] (0,0) (0,\the\dimexpr\fontdimen22\textfont2\relax)
  -- (2em,\the\dimexpr\fontdimen22\textfont2\relax);%
}

\begin{document}

\title{Preparation of Vibrational Quasi-Bound States of the Transition State Complex \ce{BrHBr} from the Bihalide Ion {\ce{BrHBr-}}}

\author{Luis H. Delgado G}
    \affiliation{Departamento de Química, Universidad del Valle, A.A. 25360, Cali, Colombia.}
    
\author{Carlos A. Arango}
    \affiliation{Departamento de Ciencias Químicas, Universidad Icesi, Cali, Colombia.}
    
\author{José G. López}
    \email[Correspondence email address: ]{jose.g.lopez@correounivalle.edu.co}
    \affiliation{Departamento de Química, Universidad del Valle, A.A. 25360, Cali, Colombia.}   

\date{\today} 

\begin{abstract}
Efficient strategies that allow the preparation of molecular systems in particular vibrational states are important in the application of quantum control schemes to chemical reactions. In this paper, we propose the preparation of quasi--bound vibrational states of the collinear transition state complex $\ce{BrHBr}$, from vibrational states of the bihalide ion $\ce{BrHBr-}$, that favor the bond selective breakage of $\ce{BrHBr}$. The results shown complement the investigation that we reported in a previous paper, [A. J. Garzón-Ramírez, J. G. López and C. A. Arango, \textit{Int.  J.  Quantum  Chem.}, 2018, \textbf{24}, e25784], in which we demonstrated the feasibility of controlling the bond selective decomposition of the collinear BrHBr using linear combinations of reactive resonances. We employed a dipole moment surface, calculated at the QCISD/\textit{d-aug-cc-pVTZ} level of theory, to simulate the interaction of the $\ce{BrHBr-}$ ground vibrational state with heuristically optimized sequences of ultrashort infrared linear chirped laser pulses to achieve a target vibrational state, resulting from expanding a chosen linear combination of reactive resonances of BrHBr in terms of vibrational eigenstates of $\ce{BrHBr-}$. The results of our simulations show final states that capture
 the most relevant features of the target state with different levels of description depending on the sequence of laser pulses employed. We also discuss ways of improving the description of the target state and possible limitations of our approach.
\end{abstract}

\maketitle

\section{Introduction} \label{sec:outline}

Controlling the pathways and enhancing the yield of specific products in a chemical reaction is of central importance in chemical research.\cite{brumer1989,shapiro2003,gordon1999} Among the ways of controlling the fate of a chemical process, quantum coherent control of molecular processes provides the means to guide and control a chemical reaction towards specific products, with a selectivity that is not possible to obtain with the most common chemical and photochemical methods.\cite{rabitz2000} The selectivity exhibited by quantum coherent control depends mainly on the preparation of a target molecular state, which is achieved using multiple interfering pathways enabled by the interaction of laser light with the molecule. \cite{brumer2001}

Among the types of lasers, in terms of the duration of the laser emission, ultrashort shaped laser pulses have been successful in the experimental implementation of laser-induced control of molecular systems.\cite{lozovoy2008} This is due to their versatility to facilitate the necessary interferences required in the preparation of the molecular states. Some examples of the applications of ultrashort shaped laser pulses to control chemical systems are control of the yield of $\ce{H3+}$  from methanol,\cite{dantus2019} dissociative ionization of molecules following interaction with intense laser pulses,\cite{dantus2016} laser-controlled chemistry for the relative yield of product ions, and laser control on the fragmentation of n-propyl benzene. \cite{goswami2009}

One way of controlling a molecular process is through the manipulation of resonance states due to the important role they play in a variety of molecular events.\cite{seideman1999, brumer2013, garcia2017, garcia2019, garcia2019a, garcia2020} For a chemical reaction, reactive resonances can be formed as the reaction occurs.\cite{liu2012} Within the framework of the Born-Oppenheimer approximation, a chemical reaction takes place on a potential energy surface (PES) over which the nuclei evolve in time on their course from reactants to products. A region of particular interest in a PES is the transition state (TS) region where chemical bonds are broken and formed.\cite{liu2012, continetti2017, neumark2018} Due to the transient nature of a TS which involves the formation of quasi-bound states, reactive resonance states can be associated to a chemical reaction.\cite{liu2012} These states can markedly affect the course of a bimolecular encounter, reason why they are of central importance for the manipulation of the fate of a chemical process using quantum coherent control.\cite{liu2016, guo2020}

From the experimental point of view, one of the tools applied to the investigation of reactive resonances in the TS region is negative ion photoelectron spectroscopy. In this technique, a stable anion is employed as the precursor of a neutral TS complex. If the anion is similar in geometry to the TS complex, the photo-detachment of the electron places the system in the neutral TS region providing a good Franck--Condon overlap. Measurement of the kinetic energy of the ejected electron, after detachment, gives information of the vibrational states of  the TS complex and existing TS reactive resonances.\cite{continetti2017} Negative ion photoelectron spectroscopy was pioneered by the work of Neumark and his research group, who applied this technique on the triatomic system,$\ce{XHY-}$ where X and Y are halogen atoms (F, Cl, Br, and I) .\cite{neumark1988, neumark1988a, neumark1990} The corresponding unstable molecular complex $\ce{XHY}$, reached by  negative ion photoelectron spectroscopy, represents the TS of the hydrogen transfer reaction $\ce{X + HY -> XH + Y}$. Analysis of the photo-detachment spectrum of $\ce{IHI-}$ revealed the presence of reactive resonances in these systems.\cite{neumark1990a} Since then, several studies have been done to provide further physical insight into the structure of the transient complex $\ce{XHY}$ and the dynamics of the corresponding bimolecular encounter $\ce{X + HY -> XH + Y}$.\cite{neumark1992, kubach1994, morokuma2000, miller2005, takayanagi2014, takayanagi2015, takayanagi2017, mccoy2017} 

The existence of quasi--bound vibrational states in the XHY complexes, that are associated with reactive resonances, makes them amenable to propose control strategies for manipulating reactions of the type $\ce{X + HY -> XH + Y}$. In this line of research, Manz and co-workers proposed the use of reactive resonances of FDBr  to selectively enhance or suppress the hydrogen type particle transfer in the collinear $\ce{F + DBr -> FD + Br}$.\cite{bisseling1988} In a previous work, hereafter referred as paper I, we suggested the feasibility of collinear bond selective decomposition of BrHBr via linear combinations of quasi-bound vibrational states of $\ce{BrHBr}$ interacting with an ultrashort infrared (IR) linear chirped pulse.\cite{garzon2018} Another suggested way of controlling these kinds of reactions, in the line of the negative ion photoelectron experiments, employs an ultrashort few-cycle IR laser pulse to break the symmetry of the bihalide ion $\ce{FHF-}$, by displacing the ionic complex from its equilibrium position, towards a favorable configuration for the anti--symmetric breakage of the FHF complex, which is achieved with a well-timed ultrashort ultraviolet (UV) pulse. \cite{elghobashi2003breaking, elghobashi2004}

Although the results of our previous investigation suggest that controlling the dissociation of the short lived $\ce{BrHBr}$ complex by using vibrational reactive resonances is possible, remains the issue of how to prepare the system into those states that favor the bond selective breakage of $\ce{BrHBr}$.\cite{garzon2018} In the present study, we propose the preparation of specific quasi-bound vibrational states of $\ce{BrHBr}$ from vibrational states of the bihalide ion $\ce{BrHBr-}$. In particular, we use superpositions of ultrashort IR  linear chirped pulses to drive the ground vibrational state of $\ce{BrHBr-}$ towards a target vibrational state of $\ce{BrHBr-}$, which is the result of expanding a chosen linear combination of reactive resonances of $\ce{BrHBr}$ in terms of vibrational eigenstates of the corresponding bihalide ion. With the results of this work, we complement the investigation reported by us in paper I for the bond selective decomposition of $\ce{BrHBr}$ and hope to gain physical insight into the interaction of linear sequences of pulses with molecules of the type $\ce{XHY-}$. This will allow us to develop appropriate strategies required in the investigation of the manipulation of resonance states to control the course of chemical reactions.

The rest of the paper is organized as follows: In Section 2.1 we define the coordinate system, describe the Hamiltonian operator, and establish the expansion basis to construct the target state. In Section 2.2 we introduce the control strategy to drive the system from the initial ground vibrational state to the target state. In Section 3 we present and discuss the results of achieving the target state. In Section 4 we summarize the results and present the main conclusions of our proposed control strategy.

\section{Theoretical Model and Methods of Calculation} \label{sec:develop}

\subsection{Target vibrational state}

\subsubsection{Coordinates and Hamiltonian operator}

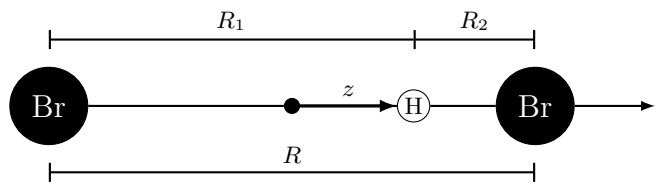
\begin{figure}[h]
    \centering
    \begin{tikzpicture}[scale=0.8]
        \node[circle,draw,text=white,fill=black, inner sep=5pt] (br1) at (0,0){\large{$\ce{Br}$}};
        \node[circle,draw,text=black,fill=white, inner sep=1pt] (h) at (6,0){$\ce{H}$};
        \node[circle,draw,text=white,fill=black,inner sep=5pt] (br2) at (8,0){\large{$\ce{Br}$}};
        \node[circle, draw,fill=black,inner sep=-2pt] (coc) at (4,0){};
        \node[circle, draw=white,fill=white,inner sep=0pt] (rightest) at (10,0){};
        \draw[black,thick] (br1.east)--(coc.west);
        \draw[black,thick] (coc.east)--(h.west);
        \draw[black,thick] (h.east)--(br2.west);
        \draw[-latex,black,thick] (br2.east)--(rightest.west);
        \draw[-latex,black,very thick] (coc.east)--node[midway, above]{$z$} (h.west);
        \draw[|-|, black,thick] (0,-1.1)--node[midway, above]{$R$} (8,-1.1);
        \draw[|-, black,thick] (0,1.1)--node[midway, above]{$R_1$} (6,1.1);
        \draw[|-|, black,thick] (6,1.1)--node[midway, above]{$R_2$} (8,1.1);
\end{tikzpicture}
    \caption{Coordinate system for the collinear $\ce{BrHBr-}$.}
    \label{fig:geometria}
\end{figure}

The target vibrational state is obtained by means of the expansion of the $\ce{BrHBr}$ quasi-vibrational state as a linear combination of vibrational eigenstates of the bihalide ion $\ce{BrHBr-}$. In this work, we restrict the geometry of $\ce{BrHBr-}$ to be collinear with the bromine nuclei aligned along the $z-$axis and the hydrogen moving collinearly between the two heavy atoms, see Figure \ref{fig:geometria}. We employ Cartesian coordinates, with the geometrical center of $\ce{Br-Br}$ as the origin of the coordinates, and atomic units in all the calculations. The Hamiltonian operator for the ionic complex $\ce{BrHBr-}$ is

\begin{equation}
\hat{H}_\text{mol}=-\frac{1}{2 \mu_{z}}\frac{\partial^2}{\partial z^2}-\frac{1}{2 \mu_{R}}\frac{\partial^2}{\partial R^2}+\hat{V}(z,R),\label{eq:hamiltonian}
\end{equation} 

\noindent where $z$ represents the position of the hydrogen atom, $R$ is the $\ce{Br-Br}$ distance, $\mu_{z}$ is the reduced mass of $\ce{[H-Br2]^-}$, and $\mu_{R}$ provides the reduced mass of the $\ce{Br-Br}$ diatom. The $\hat{V}(z,R)$  operator is the PES of $\ce{BrHBr-}$. $\hat{H}_\text{mol}$ is employed in both the calculation of the $\ce{BrHBr-}$ vibrational eigenstates and the simulation of the photoexcitation of the $\ce{BrHBr-}$ vibrational ground state.

\subsubsection{Potential energy surface}
\label{subsubsection:pes}

\begin{figure*}[ht]
    \centering
    \includegraphics[width=0.8\textwidth]{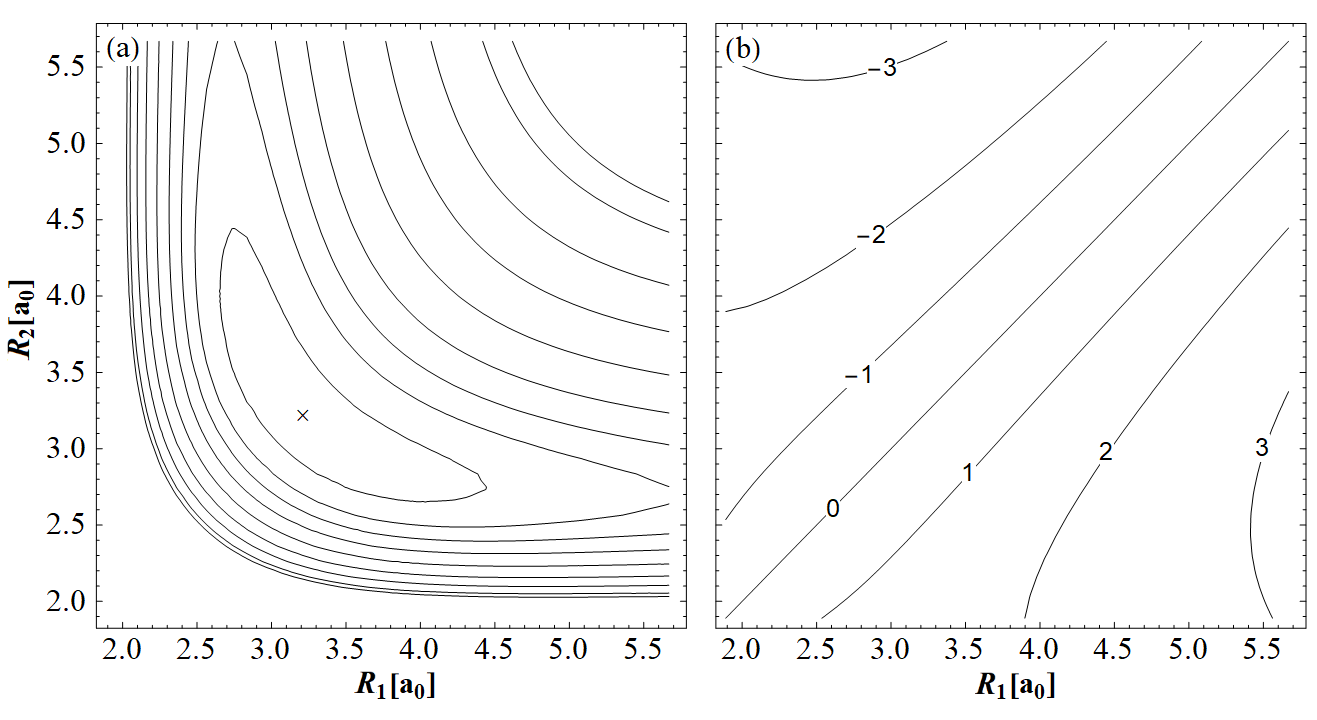}

     \caption{(a) \textit{Ab initio} potential energy surface for the collinear $\ce{BrHBr-}$. The global minimum, marked with an $\times$, is located at $R_\text{Br--Br} = 6.42$ $a_0$. The contours $0.1197, 0.3918, 0.6640, 1.0100, 1.4803, 2.0245, 2.5687, 3.1130,$ and, $3.3581$ eV are relative to the global minimum of the PES. (b) Calculated electric dipole moment surface for collinear $\ce{BrHBr-}$. The contours depicted are in D.}
    \label{fig:sep-smd}
\end{figure*}

The \textit{ab initio} PES for the ion $\ce{BrHBr-}$ was constructed from single point energy calculations at the QCISD/aug-cc-pVTZ level of theory using the software program GAUSSIAN 09. \cite{g09short} We chose this level of theory due to the good agreement with experimental results for different bihalide ions $\ce{XHX-}$, shown in previous studies. \cite{elghobashi2006} The collinear PES was computed in terms of $R_1$ and $R_2$, which are the $\ce{Br^{\text{(left)}}-H}$ and $\ce{H-Br^{\text{(right)}}}$ distances, respectively, see Figure \ref{fig:geometria}. These coordinates are related to the $(z,R)$ coordinates through 
\begin{align}
    R &= R_1+R_2, \nonumber \\ 
    z &= \frac{R_{1}-R_{2}}{2}. \label{eq:coordinatetransformationZR}
\end{align}

A grid of $41\times 41$ \textit{ab initio} points were calculated, with values of both $R_1$ and $R_2$ ranging from 1.0 to 3.0\,\AA\ and taking advantage of the symmetry of the system in which only geometries with $R_1 \ge R_2$ were used. Figure \ref{fig:sep-smd}(a) shows the computed \textit{ab initio} PES. This PES exhibits  a global minimum at the $\ce{Br-Br}$ distance of 6.42 $a_0$, which is equal to the value reported by Manz y coworkers using a highly correlated configuration interaction method,\cite{manz2015} and a shallow well, a characteristic feature when the geometry of these ionic complexes is restricted to be collinear. \cite{yamashita1993, jiang1974}

To facilitate the calculations involved, the PES was fitted using Wolfram Mathematica 11.2\cite{mathematica} with a LEPS type potential\cite{sato1955} as a fitting function. The fitted PES correctly describes the shallow well of the \textit{ab initio} PES with only a small difference in the global minimum: $R_{\ce{Br-Br}} = 6.35$ $a_0$ and $E_ \text{min}=-5146.1260$\,$E_\mathrm{h}$ for the fitted PES versus $R_{\ce{Br-Br}} = 6.42$ $a_0$ and $E_ \text{min}=-5146.1244$\,$E_\mathrm{h}$ for the \textit{ab initio} PES.

\subsubsection{\ce{BrHBr-} eigenfunctions and construction of the target vibrational state}

The $\ce{BrHBr-}$ vibrational eigenfunctions $\Psi^\text{ion}_{v_{z},v_{R}}(z,R)$ and eigenenergies $E_{v_{z},v_{R}}$, are obtained via the solution of the time independent Schrödinger equation with the fitted PES described above. We employed a discrete variable representation (DVR) scheme\cite{colbert1992} on a uniform grid of $251\times251$ points, with values of $z$ and $R$ ranging from $-2.0$ to $2.0$ $a_0$ and $5.0$ to $10.0$ $a_0$, respectively.

The quasi-bound vibrational state of BrHBr employed in our calculations is the result of a linear combination of the reactive resonances $\Psi_{2,1}$ y $\Psi_{4,0}$ of $\ce{BrHBr}$:

\begin{equation}
\Psi_\text{BrHBr}(z,R)= \sin\left(\eta\right) \Psi_{2,1} +\exp\left(i\Delta\xi\right)\cos\left(\eta\right) \Psi_{4,0},\label{eq:nwf}
\end{equation}

\noindent with  $\eta=0.33 \pi$ y $\Delta\xi=1.78 \pi$ (see paper I for details on how this vibrational state was constructed). We chose this combination since it exhibits the largest displacement of the initial wave packet $\Psi_{\ce{BrHBr}}(z,R)$ towards the dissociation channel with $z<0$ as reported in paper I. We express this state in terms of an expansion of the vibrational eigenstates of $\ce{BrHBr-}$:\cite{levine2014}

\begin{equation}
\Psi_\text{BrHBr}(z,R)= \sum_{v_{z}} \sum_{v_{R}} c_{v_{z},v_{R}} \Psi_{v_{z},v_{R}}^\text{ion}(z,R),\label{eq:expansion}
\end{equation}

\noindent with values of the quantum number $v_z$ y $v_R$ ranging from 0 to 4. The expansion coefficients $c_{v_{z},v_{R}}$ are given by:

\begin{equation}
c_{v_{z},v_{R}} = \int_{0}^{\infty} \int_{-R}^{R} \Psi_{v_{z},v_{R}}^\text{ion}(z,R')^{*}\Psi_\text{BrHBr}(z,R')\,dz dR'.\label{eq:coe}
\end{equation} 

The expansion \eqref{eq:expansion} is the target state. The coefficients were calculated using Wolfram Mathematica 11.2.\cite{mathematica} 

\subsection{Target state preparation by means of laser pulses}

\subsubsection{Dynamics of \ce{BrHBr-} by photo-excitation}

 Once the target state is constructed using Equations \ref{eq:expansion} and \ref{eq:coe}, we photo-excite the ground vibrational state of the collinear $\ce{BrHBr-}$ in its lowest electronic state, via IR laser pulses, in order to drive the initial state to the target state. The simulation of the dynamics of $\ce{BrHBr-}$ is based on the solution of the time dependent Schrödinger equation (TDSE), 
 
\begin{equation}
 \textit{i}\frac{\partial \Psi(t)}{\partial t} = \hat{H}(t) \Psi(t), \label{eq:STDE}
\end{equation}

\noindent where $\Psi(t)=\Psi(z,R,t)$ represents the $\ce{BrHBr-}$ wave function at the time \textit{t}. $\hat{H}(t)$ is the time-dependent Hamiltonian operator
\begin{equation}
\hat{H}(t)=\hat{H}_\text{mol}+ \hat{H}_\text{int}(t). \label{eq:mol+int} 
\end{equation}

\noindent with $\hat{H}_\text{mol}$ as the Hamiltonian operator of $\ce{BrHBr-}$, described in Equation \ref{eq:hamiltonian}, and $\hat{H}_\text{int}(t)$ represents the interaction of the molecule with the laser field. In this work, we approximate $\hat{H}_\text{int}(t)$ by
\begin{equation}
\hat{H}_\text{int}(t)=-\hat{\mu}(z,R)\varepsilon(t), \label{eq:inthamiltonian} 
\end{equation}

\noindent with $\hat{\mu}(z,R)$ is the electric dipole moment function of the molecule and $\varepsilon(t)$ is the laser field.

We employed the interaction picture of quantum mechanics to solve the TDSE for $\Psi(t)$, \cite{merzbacher1998} in which

\begin{equation}
 \textit{i}\frac{\partial \widetilde{\Psi}(t)}{\partial t} = \widetilde{H}_\text{int}(t) \widetilde{\Psi}(t). \label{eq:NEWSTDE}
\end{equation}

$\widetilde{\Psi}(t)$ is given by

\begin{equation}
\widetilde{\Psi}(t)=\exp( \textit{i} \hat{H}_\text{mol}t)\Psi(t), \label{eq:NEWwf}
\end{equation}

\noindent and the $\widetilde{H}_\text{int}(t)$ operator is
\begin{equation}
\widetilde{H}_\text{int}(t) = \exp\left(i \hat{H}_\text{mol}t\right)  \hat{H}_\text{int}(t) \exp\left(- i\hat{H}_\text{mol}t\right). \label{eq:NEWhamiltonian}
\end{equation}

$\widetilde{\Psi}(t)$ is the target state, Equation \ref{eq:expansion}, with unknown time-dependent coefficients $c_{v_{z},v_{R}}(t)$

\begin{equation}
\widetilde{\Psi}(t)= \sum_{v_{z}=0}^{\infty} \sum_{v_{R}=0}^{\infty} c_{v_{z},v_{R}}(t) \Psi_{v_{z},v_{R}}^\text{ion}.\label{eq:NEWexpansion}
\end{equation}

The $c_{v_{z},v_{R}}(t)$ coefficients were obtained by solving the following set of coupled differential equations:

\begin{equation}
\textit{i} \frac{d{c}_{v_{z}',v_{R}'}(t)}{dt}  =\sum_{v_{z}=0}^{\infty} \sum_{v_{R}=0}^{\infty}e^{\textit{i} \omega_{v_z',v_R',v_z,v_R}t} H_{\text{int},v_z',v_R',v_z,v_R}(t)  c_{v_{z},v_{R}}(t),\label{eq:expansion+TDSE}
\end{equation}

\noindent where
\begin{equation}
H_{\text{int},v_{z}',v_{R}',v_{z},v_{R}} (t)= \int_{- \infty}^{\infty} \int_{- R}^{R} [\Psi_{v_{z}',v_{R}'}^\text{ion}(z,R')]^{*}\hat{H}_\text{int} (t)\Psi_{v_{z},v_{R}}^\text{ion}(z,R')\, dz dR'. \label{eq:NEWintterm}
\end{equation}

In our work, $v_z', v_z, v_R',$ and $v_R$ range from 0 to 4 which give a set of 25 coupled differential equations to solve.

\subsubsection{Dipole moment surface and transition matrix}

We calculated the dipole moment surface employing the same grid points, level of theory, and computer program used in the PES calculation, see section \ref{subsubsection:pes}. Figure \ref{fig:sep-smd}(b) shows the computed dipole moment surface, which exhibits the expected symmetric growth of the magnitude of the dipole moment along the $z = (R_1-R_2)/2$ coordinate, due to the symmetry of the molecule.

An educated guess to identify favorable transitions between the $\ce{BrHBr-}$ vibrational states that could be useful to reach the target state, can be found by considering the transition matrix elements (TMEs) \cite{guerrero2018,demtroder2010}

\begin{equation}
\mu_{v_{z}',v_{R}',v_{z},v_{R}}=\int_{0}^{\infty}\int_{-R}^{R}\Psi^\text{ion}_{v_{z}',v_{R}'}(z,R')^{*}\hat{\mu}(z,R')\Psi^\text{ion}_{v_{z},v_{R}}(z,R')\,dz dR', \label{eq:MTD} 
\end{equation}

\noindent where the dipole moment operator, $\hat{\mu}(z,R)$, is the dipole moment surface described above. We calculated the TMEs with values of the quantum numbers $v_z', v_z, v_R',$ and $v_R$ ranging from 0 to 4, using Wolfram Mathematica 11.2.\cite{mathematica} To reduce the computational cost of the integration in Equation \ref{eq:MTD}, we used the same approach as in the calculations involved in the PES, that is fitting the dipole moment surface and taking advantage of the symmetry of the system.
 
\subsubsection{Laser pulses}  

In this work, we used combinations of linear chirped laser pulses (LCPs) to approximate the laser field $\varepsilon(t)$ that interacts with the $\ce{BrHBr-}$ molecule,

\begin{equation}
    \varepsilon(t) = \sum_j \varepsilon_j(t)
\end{equation}

Here $\varepsilon_j(t)$ is the $j$th LCP of the form \cite{cao1998}


\begin{equation}
\varepsilon_j(t)=\varepsilon_{0,j} \exp\left(-\frac{(t-t_{0,j})^2}{2 \tau_{j}^2}\right)\cos{\left(\omega_{0,j} (t-t_{0,j})+ \frac{c_j}{2}(t-t_{0,j})^2\right)},\label{eq:chirppulse}
\end{equation}

\noindent where $\varepsilon_{0,j}$, $\omega_{0,j}$, and $\tau_j$ represent the maximum amplitude, central frequency, and time duration of the $j$-th pulse, respectively. $t_{0,j}$ specifies the time for the $j$-th pulse highest intensity, and $c_j$ is the linear chirp constant\cite{cao1998} of the $j$-th pulse. 

We performed a heuristic search of the parameters $\varepsilon_{0,j}$, $\omega_{0,j}$, $\tau_j$, $t_{0,j}$, and $c_j$ for each $j$-th pulse in such a way that the overlap of the propagated wave function $\widetilde{\Psi}(t)$ with the target state $\Psi_\text{BrHBr}(z,R)$,

\begin{equation}
J(\widetilde{\Psi}(t))=|\widetilde{\Psi}(t)^*\Psi_\text{BrHBr}(z,R)|^2,\label{eq:NEWfitnessfunction}
\end{equation}

\noindent is maximized at the end of the laser field interaction.\cite{werschnik2007} This allows us to find the best laser pulse $\varepsilon(t)$, for different numbers of combined LCPs, that drives the $\ce{BrHBr-}$ ionic complex from its vibrational fundamental state to the target state $\Psi_\text{BrHBr}(z,R)$. The functions $\widetilde{\Psi}(t)$ and $\Psi_\text{BrHBr}(z,R)$ employed in this work, Equations \ref{eq:NEWexpansion} and \ref{eq:expansion}, respectively, lead to the following $J$ function to be maximized:

\begin{equation}
J(\widetilde{\Psi}(t))=\left|\sum_{v_z'=0}^{4} \sum_{v_R'=0}^{4} \sum_{v_z=0}^{4} \sum_{v_R=0}^{4} c_{v_z',v_R'}(t)^* c_{v_z,v_R}(t)  \delta_{v_z',v_R',v_z,v_R} \right|^2.\label{eq:finalfitnessfunction}
\end{equation}

\section{Results and Discussion} \label{sec:results}
\subsection{Target vibrational state}

\begin{figure}[ht]
    \centering
    \includegraphics[width=0.48\textwidth]{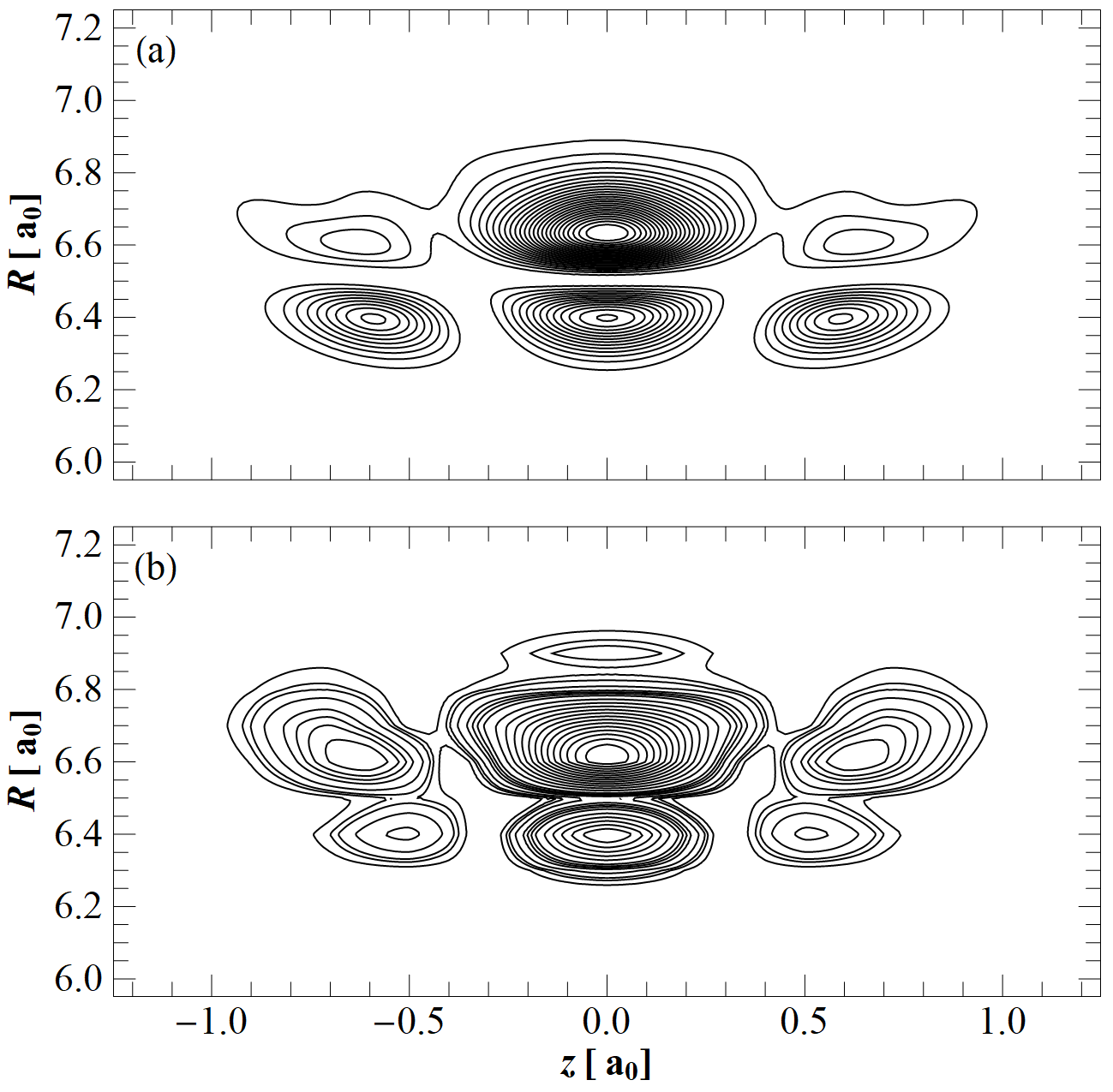}
    \caption{(a) Squared modulus of the quasi-bound vibrational state of BrHBr. (b) Squared modulus of the target state $\Psi_{\ce{BrHBr}}(z,R)$, resulting from the expansion in terms of the eigenstates of $\ce{BrHBr-}$ with $v_z$ and $v_R$ ranging from 0 to 4.}
    \label{fig:exact&expanwf}
\end{figure}

As mentioned above, the target state $\Psi_{\ce{BrHBr}}(z,R)$ is the result of the expansion of a quasi-bound vibrational state of BrHBr, Equation \ref{eq:nwf}, in terms of the vibrational eigenstates of $\ce{BrHBr-}$, Equations \ref{eq:expansion} and \ref{eq:coe}. Figure \ref{fig:exact&expanwf}(a) shows the squared modulus of the quasi-bound vibrational state of $\ce{BrHBr}$. We observe two nodes along the $z$ coordinate and one node along the $R$ coordinate. This arrangement of nodes gives six regions of high probability density, whose locations are symmetric along the $z$ coordinate, being the highest and broadest the one located around $z=0.0$ $a_0$ and $R=6.65$ $a_0$. These features are the most relevant of the quasi-bound vibrational state of interest and are captured by the constructed target state, as seen in Figure \ref{fig:exact&expanwf}(b), using values of the quantum numbers $v_z$ and $v_R$ ranging from 0 to 4. In Table \ref{tab:coef}, we report the squared modulus of the twenty five expansion coefficients $c_{v_{z},v_{R}}$ employed in the construction of the target state. Summation of these values shows a description of 84.9\%\ of the quasi-bound state by the target state.

\begin{table}[ht]
\small
\caption{\label{tab:coef} 
Squared modulus of the twenty five expansion coefficients $c_{v_{z},v_{R}}$ employed in the construction of the target state}
\begin{tabular}{ccccccccccc} \hline
 $v_z$ & $v_R$ & $|c_{v_{z},v_{R}}|^2$ & & $v_z$ & $v_R$ & $|c_{v_{z},v_{R}}|^2$ & & $v_z$ & $v_R$ & $|c_{v_{z},v_{R}}|^2$\\
\hline
0 & 0 & 0.0034 &  & 2 & 0 & 0.2212 &  & 4 & 0 & 0.0065\\
0 & 1 & 0.1615 &  & 2 & 1 & 0.0042 &  & 4 & 1 & 0.0127\\
0 & 2 & 0.1414 &  & 2 & 2 & 0.0269 &  & 4 & 2 & 0.0127\\
0 & 3 & 0.0232 &  & 2 & 3 & 0.1066 &  & 4 & 3 & 0.0107\\
0 & 4 & 0.0029 &  & 2 & 4 & 0.1068 &  & 4 & 4 & 0.0075\\
\\
1 & 0 & 0.0000 &  & 3 & 0 & 0.0000 &  &   &   &\\
1 & 1 & 0.0000 &  & 3 & 1 & 0.0000 &  &   &   &\\
1 & 2 & 0.0000 &  & 3 & 2 & 0.0000 &  &   &   &\\
1 & 3 & 0.0000 &  & 3 & 3 & 0.0000 &  &   &   &\\
1 & 4 & 0.0000 &  & 3 & 4 & 0.0000 &  &   &   &\\
\hline
\end{tabular}
\end{table}

The fact that the target state exhibits the main features of the quasi-bound vibrational state employing values of the quantum number $v_z$ and $v_R$ up to only 4, allows to test the feasibility of our approach more efficiently, due to the relatively small number of coupled differential equations, Equation \ref{eq:expansion+TDSE}, to solve. This reduces the computational cost of the calculations involved in the interaction of the initial quantum state with the laser field to drive it to the target state.

\begin{figure*}[!ht]
    \centering
    \includegraphics[width=0.8\textwidth]{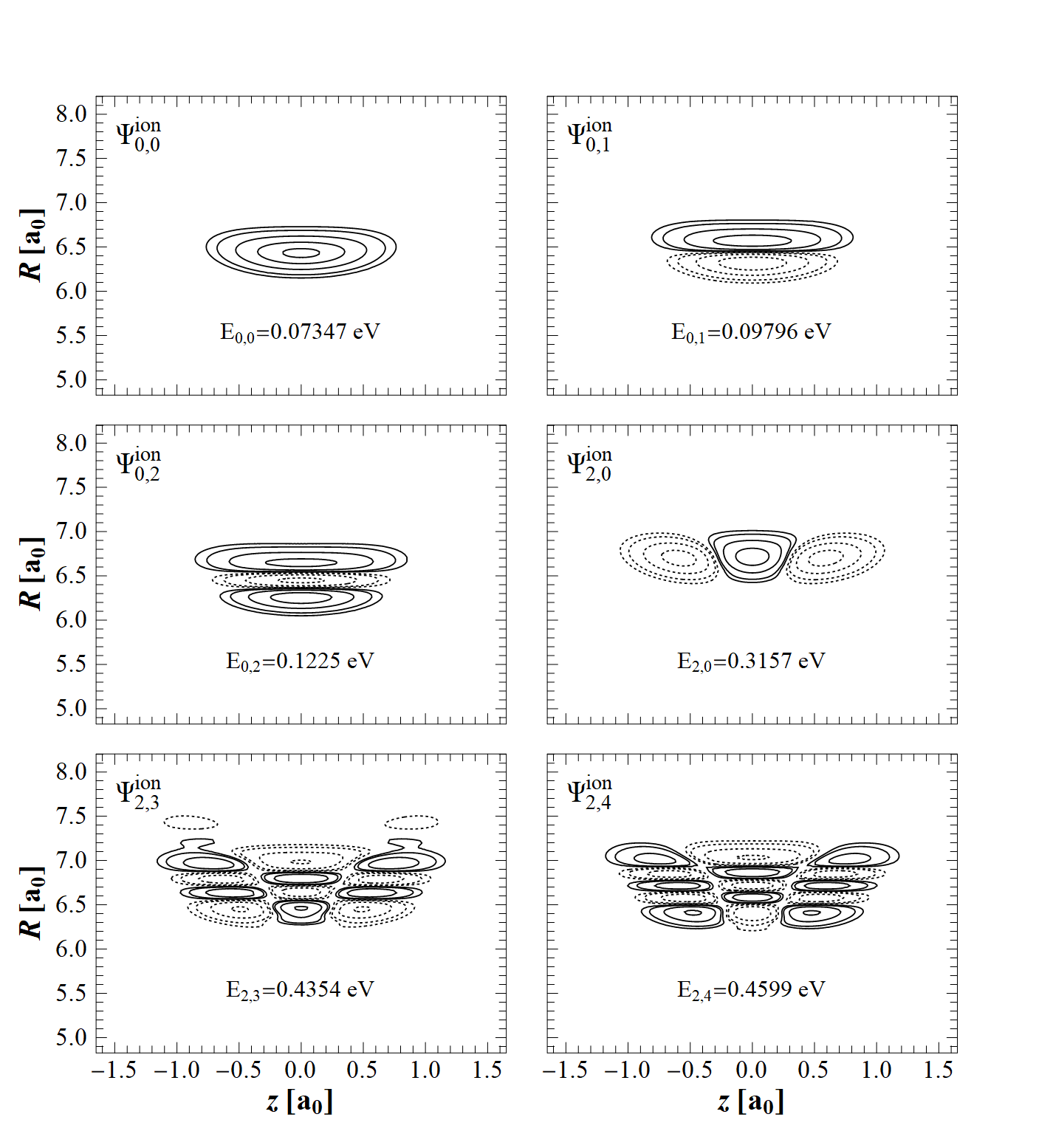}
    \caption{Vibrational eigenstates $\Psi^{\text{ion}}_{v_{z},v_{R}}(z,R)$ of the ionic complex $\ce{BrHBr-}$ with the major contribution to the constructed target state. The dashed line (\sampleline{dashed}) indicates the negative region of $\Psi^{\text{ion}}_{v_{z},v_{R}}(z,R)$, while the solid line (\sampleline{}) shows the positive region. The quantum numbers $v_z$ y $v_R$ represent the asymmetric and symmetric modes of the molecule, respectively. The energy values are relative to the global minimum of the fitted PES.}
    \label{fig:ionwf}
\end{figure*}

The values of the squared modulus of the expansion coefficients show that the eigenstates 
$\Psi^\text{ion}_{v_z,v_R}$ with even values of $v_z$ are the ones that contribute to the target state. This looks reasonable due to the symmetry of the quasi-bound state along the $z$ coordinate, see Figure \ref{fig:exact&expanwf}. Table \ref{tab:coef} also shows that the five most contributing eigenstates of $\ce{BrHBr-}$ are those with quantum numbers $(v_z,v_R) = (0,1), (0,2), (2,0), (2,3)$, and $(2,4)$. These eigestates provide the 73.7\%\ of the description of the quasi-bound state. In Figure \ref{fig:ionwf} we plot these eigenstates with their corresponding energies, including the fundamental vibrational eigenstate with $(v_z,v_R) = (0,0)$. Since the coordinates $z$ and $R$ represent the asymmetric and symmetric modes of the molecule, respectively, the quantum numbers $v_{z}$ and $v_{R}$ specify the energies of the asymmetric and symmetric stretches, respectively, and the number of nodes of the corresponding modes.

At first sight, it looks surprising that the eigenstate $\Psi^\text{ion}_{2,1}$ does not contribute as much as expected given that it has the same number of nodes as the squared modulus of the quasi-bound wave function. This result can be explained if we consider that the quasi-bound state is described by a complex function. Analysis of the shape of its Imaginary part reveals a lack of nodes along the $R$ coordinate, while its Real part exhibits the same number of nodes as its squared modulus. The eigenstate $\Psi^\text{ion}_{2,0}$, Figure \ref{fig:exact&expanwf}(d), is the one that contribute the most to the target state, which reflects the best compromise between the localization and width of the highest density probability region of the quasi-bound state and the number of nodes of the Real and Imaginary parts of its corresponding wave function.

\subsection{Analysis of the transition matrix elements}

\begin{figure}[ht]
    \centering
    \includegraphics[width=0.48\textwidth]{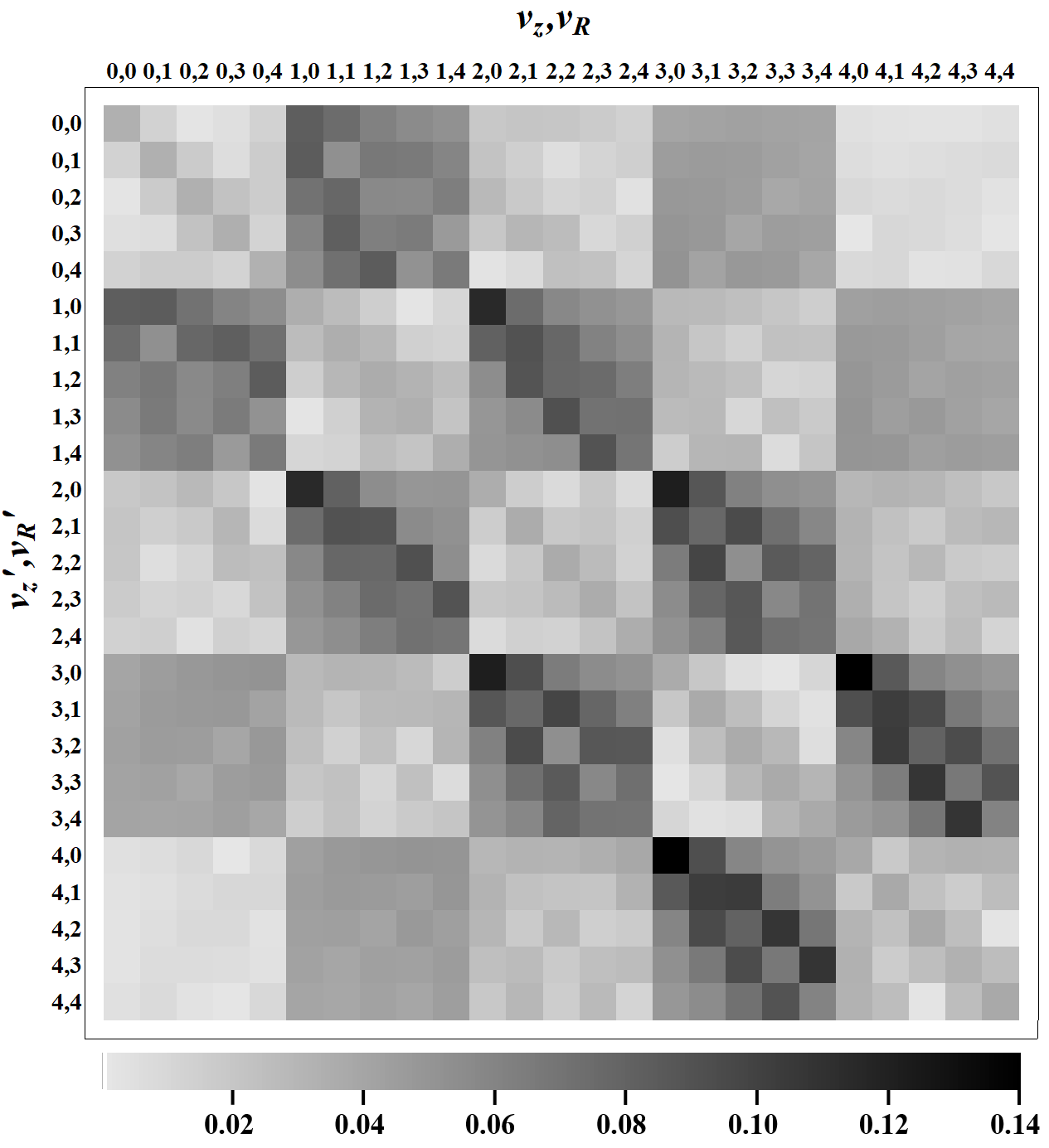}
    \caption{Squared modulus of the TMEs.}
    \label{fig:tmes}
\end{figure}

Analysis based on the TMEs allows to find an optimal pathway connecting the initial vibrational ground state of $\ce{BrHBr-}$ with the excited eigenstates of $\ce{BrHBr-}$ that contribute the most to the target state. Figure \ref{fig:tmes} shows the calculated TMEs where each small square represents the magnitude of the coupling between the states with quantum numbers $(v_z, v_R)$ and $(v_z', v_R')$. The stronger the coupling, the darker the square.

We observe that the TMEs are partitioned in a diagonal structure of squared blocks with constant values of $v_z$, in both the rows and the columns, and varying values of $v_R$. The blocks exhibiting stronger couplings are those with $\Delta v_z = v'_z - v_z$ equal to an odd value, while those with $\Delta v_z$ equal to an even value show weaker couplings. This behavior is mainly the result of the symmetry of the eigenfunctions $\Psi^\text{ion}_{v_{z},v_{R}}(z,R)$ and the dipole moment surface $\mu(z,R)$ along the $z$ coordinate. Due to the symmetry of the $\ce{BrHBr-}$ molecule, the corresponding vibrational eigenfunctions $\Psi^\text{ion}_{v_{z},v_{R}}(z,R)$ can be classified as even or odd functions along the $z$ coordinate, that is $\Psi^\text{ion}_{v_{z},v_{R}}(z,R)= \Psi^\text{ion}_{v_{z},v_{R}}(-z,R)$ for an even function (even values of $v_z$, including zero) or $-\Psi^\text{ion}_{v_{z},v_{R}}(z,R)= \Psi^\text{ion}_{v_{z},v_{R}}(-z,R)$ for an odd function (odd values of $v_z$). According to this, the dipole moment function can be classified as an odd function along the $z$ coordinate, $-\mu(z,R)=\mu(-z,R)$. Blocks with even (odd) values of $v'_z$ and odd (even) values of $v_z$ will give a resulting even product function $\Psi^\text{ion}_{v'_{z},v'_{R}}(z,R) \; \mu(z,R) \; \Psi^\text{ion}_{v_{z},v_{R}}(z,R)$ along the $z$ coordinate and, therefore, will exhibit stronger couplings between the $(v'_z,v'_R)$ and $(v_z,v_R)$ eigenstates. Blocks with even (odd) values of $v'_z$ and even (odd) values of $v_z$ will give an odd product function $\Psi^\text{ion}_{v'_{z},v'_{R}}(z,R) \; \mu(z,R) \; \Psi^\text{ion}_{v_{z},v_{R}}(z,R)$ along the $z$ coordinate, which will result in a weaker coupling between the corresponding eigenstates.

Within the blocks with strong couplings ($\Delta v_z = \pm 1$), the strongest one generally occurs when $v'_R = v_R = 0$. Given that eigenfunctions with $v_R=0$ do not have nodes along the $R$ coordinate, the product of two of them will result in a function that reflects a good overlap and hence a strong coupling is expected. Another observed feature is that the higher the $v_z$ is, the TME with $v'_R = v_R = 0$ becomes stronger. The molecular dipole moment exhibits a noticeable variation along the $z$ coordinate, which implies that eigenfunctions that are spread along this coordinate will be more affected by the variation of the dipole moment. This will cause an enhancement of the coupling between two states if they are wide enough along the $z$ coordinate. As seen in Figure \ref{fig:tmes}, less strong couplings occur between eigenstates with $\Delta v_z = \pm 1$ and $\Delta v_R = \pm 1$, followed by those with $\Delta v_z = \pm 1$ and $\Delta v_R = \pm 2$, and so on. This behavior can be explained by the increase in the number of nodes along the $R$ coordinate, as the quantum number $v_R$ increases, which will reduce the overlap between two states with increasing values of $v_R$ in one of them.

\subsection{Pulse--molecule interaction dynamics}

Our aim in this work is to photoexcite the vibrational ground state of $\ce{BrHBr-}$ in its lowest electronic state, under the interaction of a laser field, to drive the system to the target state. To accomplish this, we heuristically optimize the parameters of a composite sequence of LCPs to achieve an optimal laser pulse capable of producing a state that resemble the target state, by reaching the excited vibrational eigenstates of $\ce{BrHBr-}$ that contribute the most to the constructed target state. These eigenstates, along with their probabilities, are therefore our objective in the proposed scheme.

The TMEs reflecting stronger couplings between two eigenstates of $\ce{BrHBr-}$, Figure \ref{fig:tmes}, indicate that an optimal pathway able to connect the initial ground state with the target eigenstates must include the coupling of eigenstates such that $\Delta v_z = \pm 1$ and $\Delta v_R = 0$, as well as $\Delta v_z = \pm 1$ and $\Delta v_R = \pm 1$. In order to have a good selectivity, we employ a superposition of pulses with different values of central frequencies, each one corresponding to the energy difference between eigenstates reflecting the coupling described previously.

\begin{figure}[ht]
    \centering
    \includegraphics[width=0.48 \textwidth]{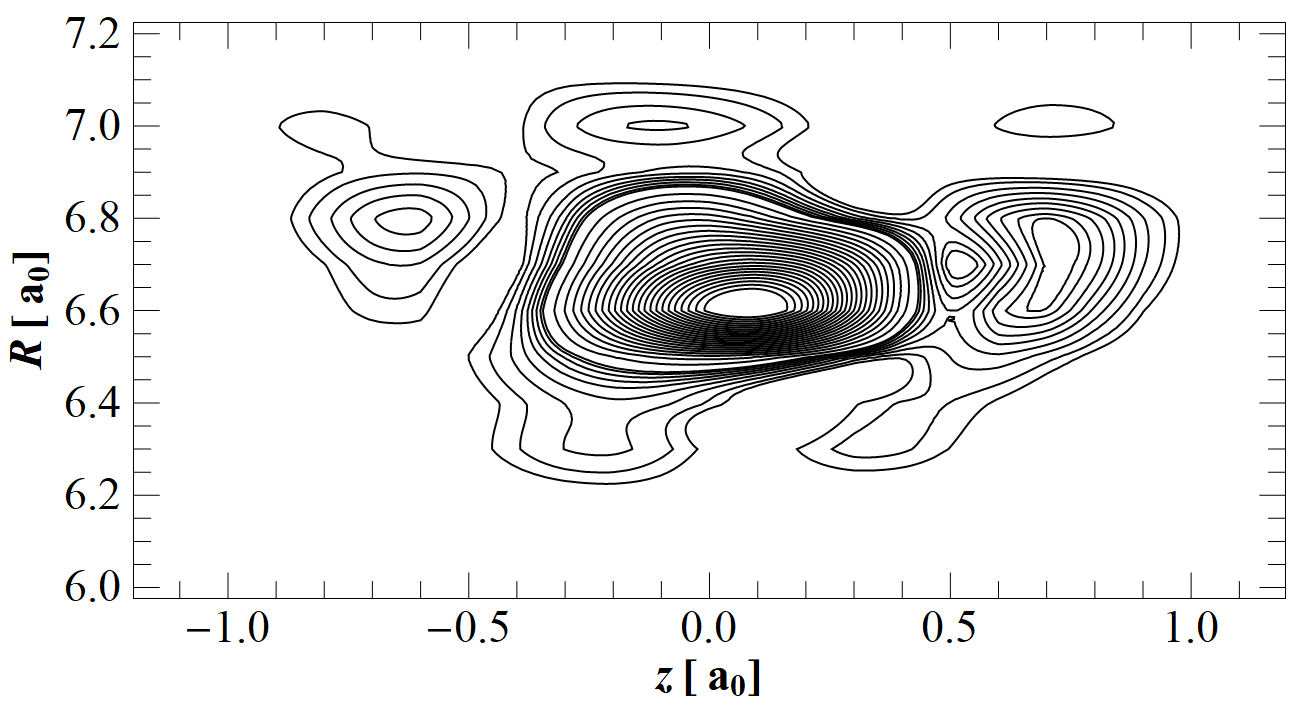}
    \caption{Squared modulus of the final state obtained under the interaction of the sequence of six LCPs, see Table \ref{tab:param-pulsos}(d).}
    \label{fig:wff}
\end{figure} 

Based on the values of the squared modulus of the expansion coefficients, we chose the states $\Psi_{2,0}^\text{ion}$, $\Psi_{0,2}^\text{ion}$, and $\Psi_{0,1}^\text{ion}$ as our target eigenstates. According to the TMEs, the coupling between the initial state $\Psi_{0,0}^\text{ion}$ with the target eigenstates is extremely weak, reason why an intermediate state needs to be considered. Analysis of the TMEs shows that the state $\Psi_{1,0}^\text{ion}$ is a good candidate to be such an intermediate state, due to the strong coupling with the initial ground state and the most contributing eigenstates to the expansion of the target state, $\Psi_{2,0}^\text{ion}$ and $\Psi_{0,1}^\text{ion}$, and therefore is included in our proposed scheme. We selected the values of $0.004154$, $0.004790$, and $0.003250$ a.u. for the central frequencies of the combined LCPs, $\omega_0$, corresponding to the energy differences $\Psi^\text{ion}_{1,0}-\Psi^\text{ion}_{0,0}$, $\Psi^\text{ion}_{2,0}-\Psi^\text{ion}_{1,0}$, and $\Psi^\text{ion}_{1,0}-\Psi^\text{ion}_{0,1}$, respectively.

It is important to consider the intensity and time duration of the laser pulse in order to avoid an undesirable ionization of the system and minimize the spontaneous relaxation of the achieved excited states. We used a value of $\varepsilon_0 = 0.003$ a.u. for the field amplitude of the pulses. This gives an intensity of the sequence of pulses below $10^{13}$ W/cm$^{2}$, which is the intensity threshold reported for these types of ionic complexes.\cite{elghobashi2004} We also restricted the time window of the simulation to the order of magnitude of $10^2$ fs, which is below the estimated lifetime of $10^4$ fs for the excited vibrational states of FHF$^-$,\cite{anderson1973} which belongs to the same bihalide series of BrHBr$^-$.

\begin{table*}[!ht]
\small
\centering
\caption{Parameters of the composite laser field sequences of LCPs employed in this work along with their percentage of description of the target state. The LCPs in each sequence of pulses are organized from lowest to highest value of $t_0$} 
\begin{tabular}{cccccccc}
\multicolumn{8}{l}{(a) Sequence of two LCPs}\\
\hline
LCP            & $\varepsilon_0$ (a.u.)  & $P_0$ (a.u.) & $\tau_j$ (a.u.) & $t_0$ (a.u.) & $\omega_0$ (a.u.) & $c$ (a.u.) & Percentage of description\\
\hline
$1$                       &                         &            &             & $10.0\times10^3$ & $0.004154$                         \\
$2$&                     
\multirow{-2}{*}{$0.003$} & \multirow{-2}{*}{$0.0095$} & \multirow{-2}{*}{$1.06\times10^{3}$} & $12.5\times10^3$ & $0.004790$ & \multirow{-2}{*}{$-2.294\times10^{-6}$} & \multirow{-2}{*}{$4,6 \%$} \\
\hline
\\
\multicolumn{8}{l}{(b) Sequence of three LCPs}\\
\hline
LCP            & $\varepsilon_0$ (a.u.)  & $P_0$ (a.u.) & $\tau_j$ (a.u.) & $t_0$ (a.u.) & $\omega_0$ (a.u.) & $c$ (a.u.) & Percentage of description \\
\hline
$1$                       &                         &          &               & $10.0\times10^3$       &$0.004154$&                  \\
$2$&                 &        &                         & $10.0\times10^3$  & $0.003250$&    \\
$3$ & \multirow{-3}{*}{$0.003$} & \multirow{-3}{*}{$0.007$} & \multirow{-3}{*}{$778$} & $14.5\times10^3$& $0.004790$& \multirow{-3}{*}{$ -1.970\times10^{-6}$}& \multirow{-3}{*}{$13,2 \%$}\\
\hline
\\
\multicolumn{8}{l}{(c) Sequence of four LCPs}\\
\hline
LCP            & $\varepsilon_0$ (a.u.)  & $P_0$ (a.u.) & $\tau_j$ (a.u.)  & $t_0$ (a.u.) & $\omega_0$ (a.u.) & $c$ (a.u.) & Percentage of description\\
\hline
$1$                       &                         &              &           & $10.0\times10^3$  & $0.004154$                       \\
$2$&                    &     &                         & $11.5\times10^3$& $0.003250$ \\
$3$                        &                         &               &          & $13.0\times10^3$ & $0.004790$                       \\
$4$ & \multirow{-4}{*}{$0.003$} & \multirow{-4}{*}{$0.0095$} & \multirow{-4}{*}{$1.06\times10^3$} & $16.0\times10^3$& $0.004790$ &\multirow{-4}{*}{$-1.667\times 10^{-6}$} &\multirow{-4}{*}{$19.7 \%$}\\
\hline
\\
\multicolumn{8}{l}{(d) Sequence of six LCPs}\\
\hline
LCP            & $\varepsilon_0$ (a.u.)  & $P_0$ (a.u.)& $\tau_j$ (a.u.)   & $t_0$ (a.u.) & $\omega_0$ (a.u.) & $c$ (a.u.) & Percentage of description\\
\hline
$1$                       &                         &             &            & $10.0\times10^3$ & $0.004154$   
\\$2$                       &                         &            &             & $10.3\times10^3$  & $0.004154$    \\
$3$&                   &      &                         & $11.5\times10^3$& $0.003250$
\\
$4$                        &                         &             &            & $12.25\times10^3$ & $0.004790$        \\
$5$                        &                         &             &            & $13.0\times10^3$ & $0.004790$                       \\
$6$ & \multirow{-6}{*}{$0.003$} & \multirow{-6}{*}{$0.0095$} & \multirow{-6}{*}{$1.06\times10^3$} & $16.0\times10^3$& $0.004790$ &\multirow{-6}{*}{$-3.033\times 10^{-7}$} &\multirow{-6}{*}{$35.8 \%$}\\
\hline
\end{tabular}
\label{tab:param-pulsos}
\end{table*}

In this work, we propose a heuristic approach to find an optimal sequence of LCPs to achieve the target state. The result of our approach is a final state that captures the most relevant features of the target state with varying levels of description depending on the number of LCPs used in the sequence of pulses. We employed several sequences of pulses, initiating from 2 up to 6 LCPs, and quantified the description of the target state by means of the squared modulus of the overlap integral,

\begin{equation}
\int_{0}^{\infty}\int_{-R}^{R}\Psi_\text{BrHBr}(z,R')^{*}\Psi^\text{ion}_{v_{z},v_{R}}(z,R';t_f)\,dzdR',
\label{eq:description}
\end{equation}   

\noindent where $\Psi^\text{ion}_{v_{z},v_{R}}(z,R;t_f)$ is the obtained final state and $\Psi_\text{BrHBr}(z,R)$ is the target state. Table \ref{tab:param-pulsos} shows the values of the percentage of description of the target state calculated by Equation \ref{eq:description} and the heuristically optimized parameters for each LCP in each sequence of laser pulses. The percentage values clearly show an increase of the description of the target state with increasing number of LCPs in the sequence of pulses. This result is expected given the fact that the use of more LCPs provides more ways of manipulating a particular sequence of laser pulses and therefore, after a judicious selection of the parameters of the LCPS, a better resemblance of the final state to the target state could be obtained.

We focus our analysis on the sequence of six LCPs given that it is the one that shows the best percentage of description of the target state, 35.8 \%. The squared modulus of the obtained final state is showed in Figure \ref{fig:wff}. Even though the final state is not symmetric along the $z$ coordinate, a feature exhibited by the target state, it captures two important features of the target state: a high and broad probability density located around $z=0.0$ $a_0$ and $R=6.6$ $a_0$ and the presence of two nodes at approximately $-0.4$ and $+0.4$ $a_0$ along the $z$ coordinate and one node at $R=6.9$ $a_0$ These results suggest that, even with a relatively small percentage of description of 35.8 \%, our proposed control scheme is able to capture the most relevant features of the target state, which is mainly achieved due to our aim of reaching the most contributing states to the target state with populations as close as possible to the squared modulus of the expansion coefficients. 

\begin{figure}[ht]
    \centering
    \includegraphics[width=0.48 \textwidth]{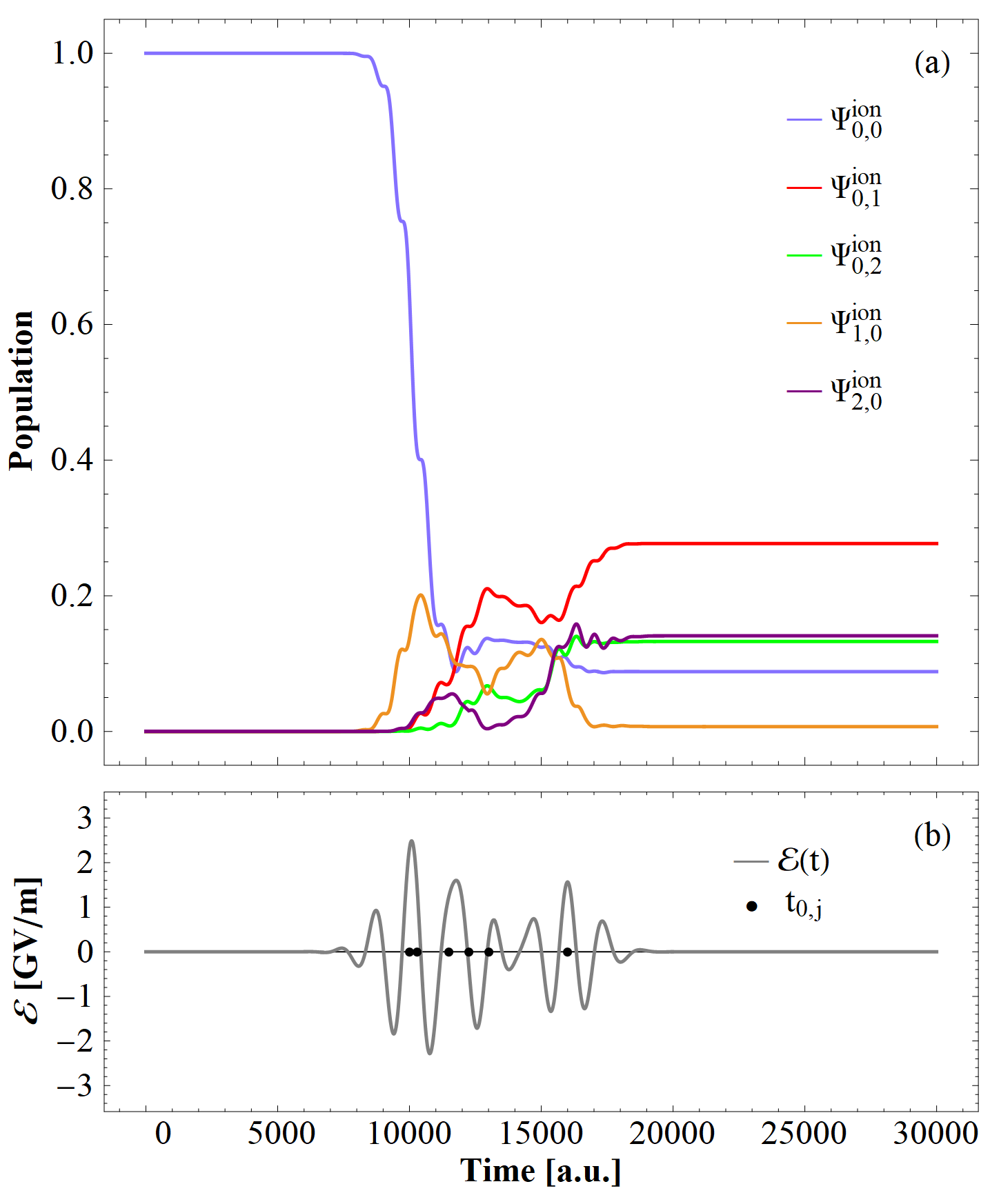}
    \caption{(a) Time evolution of the population of the vibrational eigenstates of $\ce{BrHBr-}$ $\Psi^\text{ion}_{0,0}$, $\Psi^\text{ion}_{1,0}$, $\Psi^\text{ion}_{0,1}$, $\Psi^\text{ion}_{2,0}$, and, $\Psi^\text{ion}_{0,2}$ under the interaction with a sequence of six LCPs. (b) Composite laser field sequence of six LCPs versus time. The black dots on the horizontal axis show the time of the highest intensity value for each of the six LCPs, $t_{0,j}$, as described in Table \ref{tab:param-pulsos}(d).}
    \label{fig:dynamic}
\end{figure}

The importance of the utilization of a sequence of LCPs to achieve the target state is stressed if we follow the evolution of the population of the vibrational eigenstates of $\ce{BrHBr-}$ while interacting with the the laser pulse. In figure \ref{fig:dynamic}(a) we show the populations, as function of time, of the states of interest in our study under the interaction with the sequence of 6 LCPs. This figure shows the role of the intermediate state $\Psi_{1,0}^\text{ion}$ in order to get a significant population transfer to the final states $\Psi_{2,0}^\text{ion}$, $\Psi_{0,1}^\text{ion}$, and $\Psi_{0,2}^\text{ion}$. 

Further analysis of the population of the eigenstates of Figure \ref{fig:dynamic}(a), in terms of the individual LCPs, allows to obtain a better physical insight of the effect of the sequence of laser pulses on the time evolution of the system. This is important if we want to propose new sequences of pulses that improve the description of the target state. In order to perform this analysis, we consider each of the six LCPs according to the parameter $t_{0,j}$, see Table \ref{tab:param-pulsos}(d) and Figure \ref{fig:dynamic}(b), which we take as an indicator of the time delay, with respect to the initiation of the simulation, in which each LCP might appreciably start to interact with the system and with the other LCPs. After the first and second LCP start their interaction with the molecule, we observe a transfer of population from the ground state $\Psi_{0,0}^\text{ion}$ to the intermediate state $\Psi_{1,0}^\text{ion}$, which is mainly due to the chosen central frequency of these pulses corresponding to the energy difference of these states. The third LCP, with a chosen frequency corresponding to the energy difference between the states $(1,0)$ and $(0,1)$, mainly generates a transition from $\Psi_{1,0}^\text{ion}$ to $\Psi_{0,1}^\text{ion}$. The last three LCPS are responsible for the transfer of the population from $\Psi_{1,0}^\text{ion}$ to $\Psi_{2,0}^\text{ion}$ and  $\Psi_{0,2}^\text{ion}$. 

\begin{table}[ht]
\small
\caption{\label{tab:pop} 
Final populations of the vibrational eigenstates $\Psi_{\nu_z,\nu_R}^\text{ion}$, under the interaction of a sequence of six LCPs, as described in Table \ref{tab:param-pulsos}(d)}
\begin{tabular}{ccccccccccc} \hline
 $v_z$ & $v_R$ & Population & & $v_z$ & $v_R$ & Population & & $v_z$ & $v_R$ & Population\\
\hline
0 & 0 & 0.0880 &  & 2 & 0 & 0.1409 &  & 4 & 0 & 0.0050\\
0 & 1 & 0.2769 &  & 2 & 1 & 0.0795 &  & 4 & 1 & 0.0032\\
0 & 2 & 0.1325 &  & 2 & 2 & 0.0384 &  & 4 & 2 & 0.0051\\
0 & 3 & 0.0332 &  & 2 & 3 & 0.0088 &  & 4 & 3 & 0.0043\\
0 & 4 & 0.0107 &  & 2 & 4 & 0.0135 &  & 4 & 4 & 0.0001\\
\\
1 & 0 & 0.0069 &  & 3 & 0 & 0.0431 &  &   &   &\\
1 & 1 & 0.0033 &  & 3 & 1 & 0.0385 &  &   &   &\\
1 & 2 & 0.0002 &  & 3 & 2 & 0.0176 &  &   &   &\\
1 & 3 & 0.0120 &  & 3 & 3 & 0.0087 &  &   &   &\\
1 & 4 & 0.0253 &  & 3 & 4 & 0.0041 &  &   &   &\\
\hline
\end{tabular}
\end{table}

One relevant aspect of our approach is the feasibility of improving the description of the target state. One thing we could do is manipulating the time window of some of the LCPs. For example, comparisons of the squared modulus of the expansion coefficient $|c_{0,1}|^2=0.1615$, with the corresponding population obtained in our simulations, $0.2769$ (see Table \ref{tab:pop} for the final populations of the vibrational eigenstates of $\ce{BrHBr-}$ under the interaction of six LCPs), suggest that reducing the time duration of the third pulse, the population of $\Psi_{0,1}^\text{ion}$ could be decreased. Another possibility is adding more LCPs with central frequencies corresponding to the most relevant transitions, such as $\Psi_{2,0}^\text{ion} \leftarrow \Psi_{1,0}^\text{ion}$, given that in our simulations the final population of the eigenstate $\Psi_{2,0}^\text{ion}$ is $0.1409$ versus the value of $|c_{2,0}| = 0.2212$. However, based on the way the proposed scheme is built, we consider that there is a limit, below 100 \%, in the description of the target state. This limit is imposed by the small coupling of the initial and intermediate states with the states $\Psi_{2,3}^\text{ion}$ and $\Psi_{2,4}^\text{ion}$ that have a major contribution to the target state. In order to appreciably populate those states, an approach involving successive excitations and de-excitations between states would be necessary. This would imply the use of many more LCPs in the sequence of pulses or employing sophisticated pulses that would involve the use of optimal control for their optimization, all of which would complicate the control scheme.

Once the target state is achieved, it must be driven to the PES of the neutral complex BrHBr. This step would involve a specifically designed UV laser pulse, in terms of its intensity, central frequency, time-delay with respect to the IR sequence of laser pulses employed, and time window, in order to photo-detach the electron from $\ce{BrHBr-}$. Although all of these factors have a direct impact on the excitation of the system from the ionic PES to the neutral one, the feasibility of designing an ultrashort UV pulse that drives the achieved target state, via Franck--Condon type transitions, to the neutral PES has been demonstrated in previous theoretical studies. \cite{elghobashi2003breaking,elghobashi2003,elghobashi2004,elghobashi2004breaking} 

\section{Summary and conclusions} \label{sec:conclusions}

In this work, we have proposed an approach to achieve a target state, resulting from the expansion of a quasi-bound vibrational state of the unstable complex BrHBr in terms of the vibrational eigenstates of the stable bihalide ion $\ce{BrHBr-}$. Our simulations show the feasibility of reaching specific vibrational eigenstates of the ionic complex $\ce{BrHBr-}$, through the interaction of the system with sequences of IR linear chirped laser pulses. These results suggest that, in principle, one can prepare specific quasi-bound states of the complex BrHBr with an acceptable level of description using LCPs. The success of our proposed control scheme depends on the capability of the IR sequences of LCPs of achieving the target state in the ionic PES of $\ce{BrHBr-}$ and the design of a well-timed UV pulse that drives the prepared vibrational quantum state, via photodetachment, to the neutral PES of $\ce{BrHBr}$. Once the wave packet is on the neutral PES, it would evolve with more probability towards one specific reaction channel after applying a third pulse, according to the results shown in Paper I.

Even though the highest percentage of description of the target state that we achieved was 35.8\%, a thorough analysis of the time evolution of the system suggested possible ways of improving this result, such as manipulating some of the parameters of the LCPs and/or adding more LCPs to the sequence of pulses. 
Although our results show possible limitations of the proposed scheme, due to the presence of $\ce{BrHBr-}$ eigenstates in the expanded target state that are challenging to reach, we consider that the level of description of the target state one could achieve is good enough, according to the results reported in the literature for these kinds of control approaches. \cite{Rojan2014,Gunter2021,He2021}

The results reported in this work, complement our investigation for the bond selective decomposition of the BrHBr transition state complex reported in paper I, in the sense that allow the preparation of the reported quasi-bound vibrational states of $\ce{BrHBr}$ to control the branching ratio of the product channels involved in the decomposition of $\ce{BrHBr}$: $\ce{Br + HBr}$ and $\ce{BrH + Br}$. As in paper I, the results shown assume a collinear configuration during the interaction of $\ce{BrHBr-}$ with the laser pulses, which neglects the bending modes of the molecule. Since these systems are three--dimensional, interaction of an IR pulse with $\ce{BrHBr-}$ would cause excitations of the bending modes and possibly would contribute to the overall rotation of the molecule. However, this motion should not greatly affect the directionality of the outgoing F and HF once the system is on the neutral PES.  \cite{elghobashi2003breaking} Further work taking into account these effects should be pursued. Another aspect for the consideration of the collinearity of the system under study, is the evidence that in the presence of solvating Ar atoms the decomposition of complexes of the type $\ce{XHX}$ proceeds in a nearly collinear configuration. \cite{lavender2000,neumark2001,adamovic2004,lopez2005,lopez2006} Since most of chemical reactions occur in a solvent medium, our research goes in the line of investigating control strategies of chemical reactions in solvated environments, which could be extended to other confining environments such as electrostatic cavities.

Finally, the theory of coherent control indicates that, in principle, any molecular state is achievable using laser pulses. However, not any molecular state is achieved if specific types of laser pulses and specific interactions with the molecule are employed. Our work shows that a sequence of linear chirped laser pulses interacting with the electric dipole moment of the collinear $\ce{BrHBr-}$ molecule is enough to obtain an acceptable description of the target state. For a given combination of specific types of laser pulses interacting with $\ce{BrHBr-}$ in a particular way, the issue of which would be the complete set of achievable states of the molecule is still open. We plan to address this issue in future works. 

\section*{Acknowledgements} \label{sec:acknowledgements}
This work was supported by Universidad del Valle through Project CI-71221, and by Internal Research Grants of Universidad Icesi.

\bibliography{rsc} 



\end{document}